\documentclass[sn-nature]{sn-jnl}

\usepackage{graphicx}%
\usepackage{multirow}%
\usepackage{amsmath,amssymb,amsfonts}%
\usepackage{amsthm}%
\usepackage{mathrsfs}%
\usepackage[title]{appendix}%
\usepackage{xcolor}%
\usepackage{textcomp}%
\usepackage{manyfoot}%
\usepackage{booktabs}%
\usepackage{algorithm}%
\usepackage{algorithmicx}%
\usepackage{algpseudocode}%
\usepackage{listings}%

\theoremstyle{thmstyleone}%
\theoremstyle{thmstyletwo}%

\theoremstyle{thmstylethree}%

\begin{document}

\title{Reversible ternary logic with Laguerre-Gaussian modes}


\author[1]{\fnm{Przemysław} \sur{Litwin}}

\author[1]{\fnm{Jakub} \sur{Wroński}}

\author[1]{\fnm{Konrad} \sur{Markowski}}

\author[2]{\fnm{Dorilian} \sur{Lopez-Mago}}

\author[1]{\fnm{Jan} \sur{Masajada}}

\author*[1]{\fnm{Mateusz} \sur{Szatkowski}}\email{mateusz.szatkowski@pwr.edu.pl}

\affil[1]{\orgdiv{Department of Optics and Photonics}, \orgname{Wrocław University of Science and Technology}, \orgaddress{\street{Wybrzeże Wyspiańskiego 27}, \city{Wrocław}, \postcode{50-370}, \country{Poland}}}

\affil[2]{\orgdiv{Faculty of Physics}, \orgname{University of Vienna}, \orgaddress{\street{Boltzmanngasse 5}, \city{Vienna}, \postcode{1090}, \country{Austria}}}


\abstract{The need set by a computational industry to increase processing power, while simultaneously reducing the energy consumption of data centers became a challenge for modern computational systems. In this work, we propose an optical communication solution, that could serve as a building block for future computing systems, due to its versatility. The solution arises from Landauer’s principle and utilizes reversible logic, manifested as an optical logical gate with structured light, here represented as Laguerre-Gaussian modes. We introduced an information encoding technique that employs phase shift as an information carrier and incorporates multi-valued logic in the form of a ternary system. In the experimental validation, the free space communication protocol is implemented to determine the similarity between two images. Obtained results are compared with their binary counterparts, illustrating denser information capacity and enhanced information security, which underscores its capability to transmit and process both quantum and classical information.}




\maketitle

\section{Introduction}
The recent COVID-19 pandemic accelerated the digital transformation and instantly digitized multiple life aspects on a massive scale. As has been already seen, post-pandemic life became an online-offline hybrid. Therefore, one crucial challenge for post-pandemics is increasing energy efficiency and promoting energy saving, which aligns with the Sustainable Development Goals defined by the United Nations in 2015 and summed up in 2023 \cite{UNDESA2023}.

This issue is tackled from multiple perspectives considering the ongoing technological development in various aspects. These efforts include the need to increase computer processing power \cite{Jiang2021} and provide alternative, more efficient ways to store digital data as for instance using DNA sequences \cite{Goldman2013, Matange2021}. However, as we continue proposing numerous solutions, we have to pay attention to their complementariness, so that all these efforts converge, to drive progress in a unified direction.

\begin{figure}[ht]
\centering
\includegraphics[width=8cm]{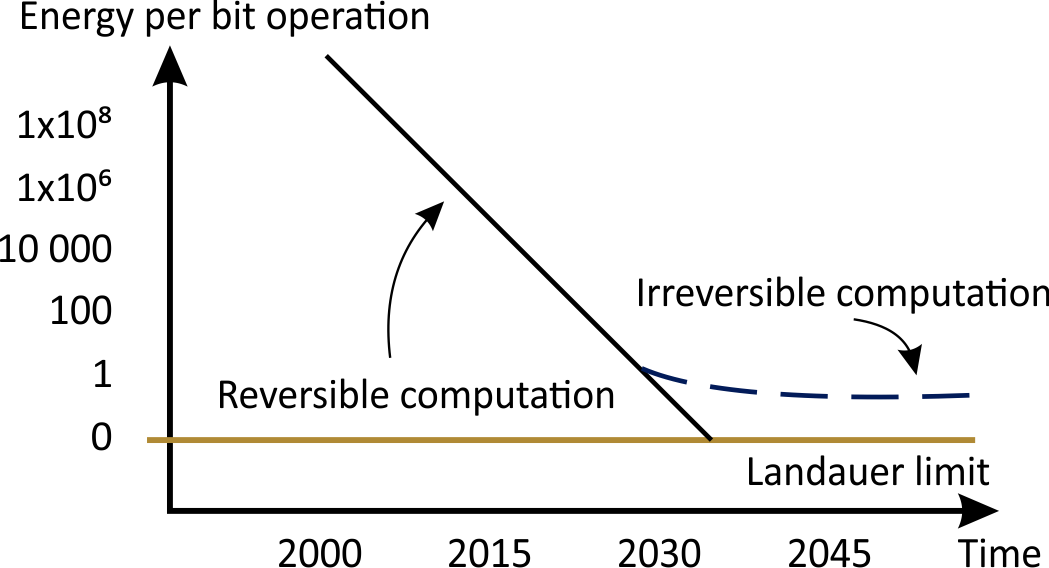}
\caption{The Landuaer's principle, comparing the energy per bit operation for both irreversible and reversible computation}
\label{fig:landauer}
\end{figure}

The proposed work focuses on developing new methods for optical communication, driven by the need set by a computational industry to continue increasing the capacity to transmit and process both quantum and classical information. One of the solutions comes up directly from Landauer's principle \cite{Landauer1961, Landauer1989} (experimentally proven in 2012 \cite{Berut2012}), which places a minimum amount of energy dissipated for each irreversible operation (i.e., bit erasure) (Figure \ref{fig:landauer} after \cite{Iniewski2019}).

The only way to reach Landauer's limit is to use reversible computation, where inputs can be reconstructed from their outputs. With computational demand on lowering the amount of energy per bit operation, we have to fundamentally change the principles we use to build computers: replace electrons with photons and implement reversible logic \cite{Hamerly2019}.

The fundamental realization of reversible logic is the controlled-SWAP gate (c-SWAP or Fredkin gate) \cite{Fredkin1982}, which consists of two information channels and one control channel that determines if a swap (exchange of information between information channels) occurs or not. The existence of only one control channel leads to numerous potential quantum information protocols including error correction \cite{Cory_1998_quantum_error_correction}, fingerprinting  \cite{Buhrman_2001_quantum_fingerprinting} \cite{Wang_2007_quantum_fingerprinting} or chip processors \cite{Li2022_fredkin_toffoli_photonics_chip}. On the other hand, one of the most promising optical communication solutions utilizes structured light, where the information is encoded in its Orbital Angular Momentum (OAM) \cite{Willner2015, Fontaine2019, Wang2022}. Based on these principles, numerous solutions have been proposed that involve not only straightforward modulation \cite{Zhu2017} but also OAM multiplexing \cite{Gong2019, Wan2022} and multicasting \cite{Fu2019}. In each of these examples, the information has been encoded in the OAM value. Recently, in contrast to these approaches, Urrego et. al. \cite{Urrego2020} proposed the c-SWAP optical gate that incorporates Laguerre-Gaussian (L-G) modes of light, where instead of the OAM value, the information has been encoded in the phase shift of the L-G mode, bringing the additional security into the communication protocol, where typical decoding methods based on the OAM content do not work \cite{Kai2019}. 

In \cite{Szatkowski2021} authors modified the phase-shifting encoding/decoding algorithm and incorporated a faster modulation device. They overcame the necessity of bit-by-bit comparison, therefore determining the similarity of two signals propagating in the free space with kHz frequency. These two concepts successfully implemented another approach to optical communication with structured light, where information is encoded into a phase-shift, instead of the OAM value. However, to level the proposed reversible computation solution up to a reliable communication system, there is still a need to increase its computation capacity.

The promising further development is to implement multi-valued logic, which, where optical protocols, in contrast to typical electrical transistors, do not require on/off states and binary data representation \cite{Athale2016, Sasikala2018}, It has already been shown that multi-valued logic can efficiently be applied to various optical computation solutions \cite{Imai1987, Jin2005, Sun2007, Zhou2009, Mourgias-Alexandris2018}. 

Such an aspect would parallel the fast speed benefits brought by the implementation of photons instead of electrons. Moreover, this could be a direct link to digital storage in DNA sequences, where multi-valued logic is utilized to convert data from their digital to nucleotide representation \cite{osti_1619517}.

In this work, we followed the phase-shifting encoding approach presented in \cite{Szatkowski2021}, but instead of two, we used three distinguished states to encode the signal, therefore successfully equipping the communication protocol with ternary logic. We show that this approach not only increases the computational capacity of the protocol but also makes the whole system more secure - one signal cannot be directly deducted even if the other signal is known.

In the experimental validation, preceded by an analytical study on the possible multi-valued solutions, we compare the performance of the ternary protocol with its binary version, showing that the protocol capacity has been significantly increased. 

While the proposed solution utilizes classical light, due to its experimental fidelity, it can be straightforwardly extended to the quantum regime, inspiring quantum applications in future scenarios and transforming modern data processing.

\section{Results}\label{sec2}
\subsection*{Ternary message-passing protocol}

In the proposed message-passing protocol, the Laguerre-Gaussian (LG) modes will act as message carriers. In the cylindrical coordinate system $(r,\phi,z)$ the complex amplitude of the $LG$ mode of topological charge $m$ can be expressed as (assuming $z=0$):
\begin{equation}
    \mathrm{LG}_m (r,\phi) = C_m \left(\frac{r}{w_0}\right)^{|m|} \exp\left( -\frac{r^2}{w_0^2} \right)\exp(i m \phi).
\end{equation}
where $w_0$ denotes the beam waist and $C_m$ is the normalizing constant. 

Following the analytical expressions presented in \cite{Urrego2020} and \cite{Szatkowski2021}, two messages are encoded in the phase shift $\alpha$ and $\beta$: $U_\alpha=A exp(i\alpha)$ and $U_\beta=B exp(i\beta)$ with amplitudes $A$ and $B$, acting on the LG modes having the same topological charge, but opposite handedness: $m_1=1$ and $m_2=-1$, respectively. Therefore, the field that enters the C-SWAP circuit having the horizontal polarization $\hat{x}$ can be expressed as:

\begin{eqnarray}
    \mathbf{E}(r,\phi)&=& 
    \left[U_\alpha LG_1(r,\phi)+U_\beta LG_{-1}(r,\phi)\right]\mathbf{\hat{x}}. 
\end{eqnarray}

Simplified as:

\begin{eqnarray}
    \mathbf{E}&=& 
    \left[U_\alpha LG_1+U_\beta LG_{-1}\right]\mathbf{\hat{x}}. 
\end{eqnarray}

The incoming field passes through the Hadamard gate, realized through the halfwave plate that sets the diagonal polarization: 

\begin{eqnarray}
    \mathbf{E}&=& 
    \left[U_\alpha LG_1+U_\beta LG_{-1}\right]\frac{\mathbf{\hat{x}}+\mathbf{\hat{y}}}{\sqrt{2}} 
\end{eqnarray}
where $\hat{y}$ denotes the vertical polarization.

Further, the field enters the c-SWAP gate, performed by an unbalanced polarization Mach-Zehnder interferometer that splits it into separate polarization components and reverses the sign of L-G modes in one of them, therefore directly implementing the swap operation:

\begin{eqnarray}
    \mathbf{E}&=& 
    \left[U_\alpha LG_1+U_\beta LG_{-1}\right]\frac{\mathbf{\hat{x}}}{\sqrt{2}}
    +[U_\alpha LG_{-1}+U_\beta LG_1]\frac{\mathbf{\hat{y}}}{\sqrt{2}}
    \label{Eq:input1}
\end{eqnarray}

Which is followed by the second Hadamard gate:

\begin{eqnarray}
    \mathbf{E}&=&
    (U_\alpha+U_\beta)(LG_1+LG_{-1})\frac{\mathbf{\hat{x}}}{2}
     +(U_\alpha-U_\beta)(LG_1-LG_{-1})\frac{\mathbf{\hat{y}}}{2}
    \label{Eq:input2}
\end{eqnarray}

Each polarization component can be separately detected through two power meters $P_x$ and $P_y$, where the output power:

\begin{eqnarray}
    P_x &\propto& |U_\alpha+U_\beta |^{2},\nonumber \\
    P_y &\propto& |U_\alpha-U_\beta|^{2}.
    \label{Eq:outpowercomp}
\end{eqnarray}

If both encoded signals are equal, so that $\alpha=\beta$, then $U_\alpha=U_\beta$ and $P_y=0$, while $P_x=P_{\mathrm{in}}$. Where $P_{\mathrm{in}}$ represents the input power and any power losses due to propagation are omitted.

\begin{figure}[ht]
\centering
\includegraphics[width=10cm]{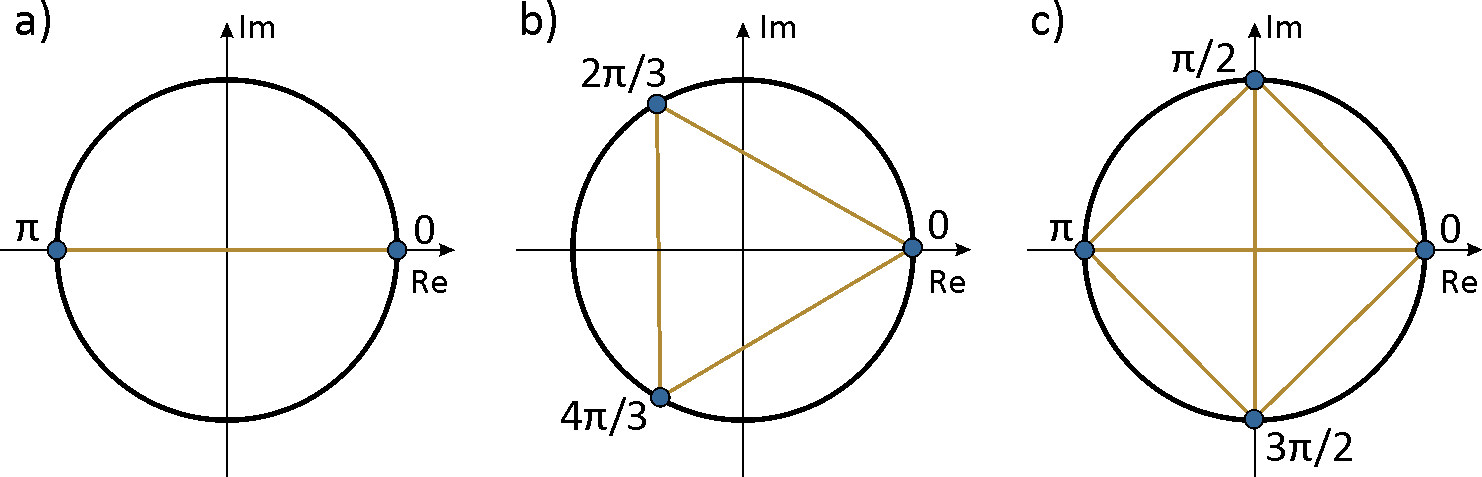}
\caption{The concept of multi-valued phase encoding, where a) presents the conventional binary encoding with [0, $\pi$] as bits [0, 1], b) ternary encoding, where [0, $2\pi/3$, $4\pi/3$] stand for trits [0, 1, 2] and c) quaternary, where [0, $\pi/2$, $\pi$, $3\pi/2$] represent quads [0, 1, 2, 3]. The blue circles indicate the chosen phase values, while the orange solid line shows the distance between all possible signal pairs.}
\label{fig:multilevel}
\end{figure}

The concept of multi-valued logic relies on the possibility of encoding the signal in either $\alpha$ or $\beta$ with the chosen phase value on the complex plane. This choice determines the smallest phase difference between subsequent elements in the preferred numerical system and it depends on the no. of values one can encode the signal through. Figure \ref{fig:multilevel} presents three examples of such numerical systems, however, denser systems can be chosen, where the increment between subsequent logical states in the phase term is expressed as $\Delta\phi_{inc}=2\pi/n$, where $n$ denotes no. of logical states in the particular numerical system ($n=2$ determines binary, $n=3$ ternary and so on). 

The denser the system, the more possible power outcomes. It strictly depend on the values of both signals that are about to be compared. In the proposed protocol, this is represented as the distance between the chosen phase states on the complex plane. As shown in Figure \ref{fig:multilevel} out of these examples, only the binary and ternary systems provide an equal distance between any possible pair. The no. of possible power levels increases with the $n$ showing an important feature of the phase encoding approach. Depending on the chosen numerical system, one can switch between boolean and fuzzy logic. The latter exists for the numerical system denser than ternary. However, to reduce the logic to boolean, one has to determine the threshold, below which the pair of signals will be considered as unequal. 

\begin{figure}[ht]
\centering
\includegraphics[width=12cm]{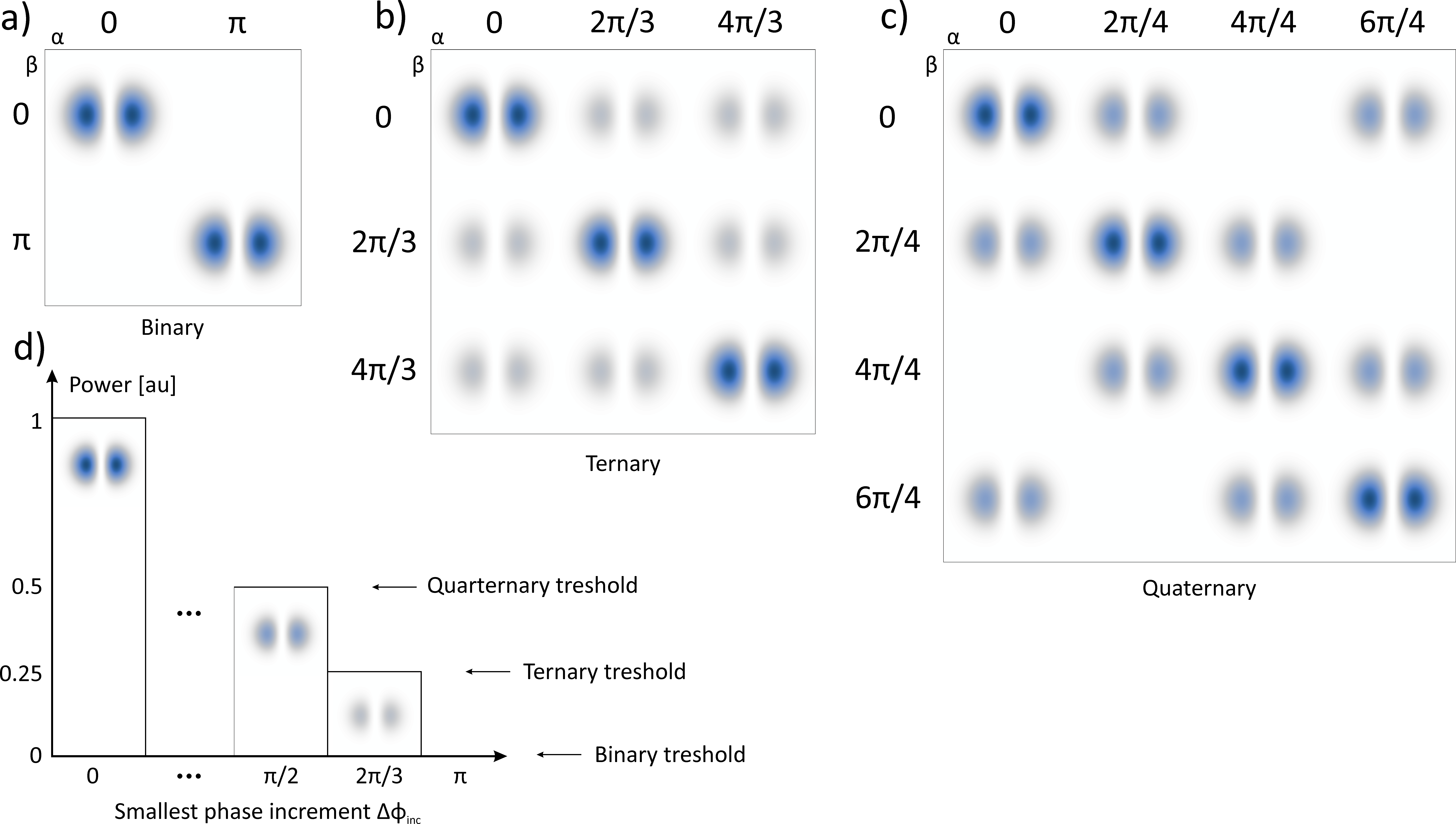}
\caption{The comparison of all available outputs of the field at the detector $P_x$ for a) binary b) ternary and c) quaternary numerical system, d) presents the possible threshold values, separating the unequal signals, when moving from fuzzy to boolean logic. All of the intensities are globally normalized.}
\label{fig:multilevel_threshold}
\end{figure}

To derive the expression for the threshold value we will consider only the detector $P_x$, as only one detector is required to determine the similarity of two signals \cite{Szatkowski2021}. The numerical simulations of the output field at the detector $P_x$ are presented in Figure \ref{fig:multilevel_threshold}a-c, where all possible combinations of $\alpha$ and $\beta$ are considered, for the chosen $n \in (2:4)$. Figure \ref{fig:multilevel_threshold}d presents the possible threshold level, determined by the increment $\Delta\phi_{inc}$, which sets the minimal distance between the neighboring logical states. The threshold value increases, as the $\phi$ gets smaller, and approaching maximum for $\phi_{inc}=0$ when both signals are equal ($\alpha=\beta$).

 If we consider the equation \ref{Eq:outpowercomp} it is visible that the power value at $P_x$ depends on the interference of two plane waves $U_\alpha$ and $U_\beta$, having the same linear polarization and perfect coherence, propagating coaxially. Thus, it is proportional to:
\begin{eqnarray}
    \mathbf{E_{s}}=U_\alpha+U_\beta
\end{eqnarray}

The intensity $I=\langle\mathbf{E_{s}}\rangle{_t}$:

\begin{eqnarray}
    I= I_1 + I_2 +2 \sqrt{I_1 I_2}cos(\Delta \Phi)
\end{eqnarray}
Where $I_1=\langle U_\alpha \rangle{_t}$, $I_2=\langle U_\beta \rangle{_t}$ and $\Delta \Phi=\alpha-\beta$. Assuming that the time-averaged intensities are equal $I_1=I_2=I_0$ and that the $I_0$ is the income intensity we get:

\begin{eqnarray}
    I= 2I_0 +2I_0cos(\Delta \Phi)
\end{eqnarray}

Thus, the output power $P_x\propto I$ relies on the phase term $\cos(\Delta\Phi)$. The boolean threshold can be determined as the $I$ value retrieved for the $\Delta \phi_{inc}$: 

\begin{eqnarray}
    I(n)= 2I_0 +2I_0cos(\Delta \phi_{inc}(n))=2I_0+2I_0cos\left(\frac{2\pi}{n}\right)
\end{eqnarray}

\begin{figure}[ht]
\centering
\includegraphics[width=5cm]{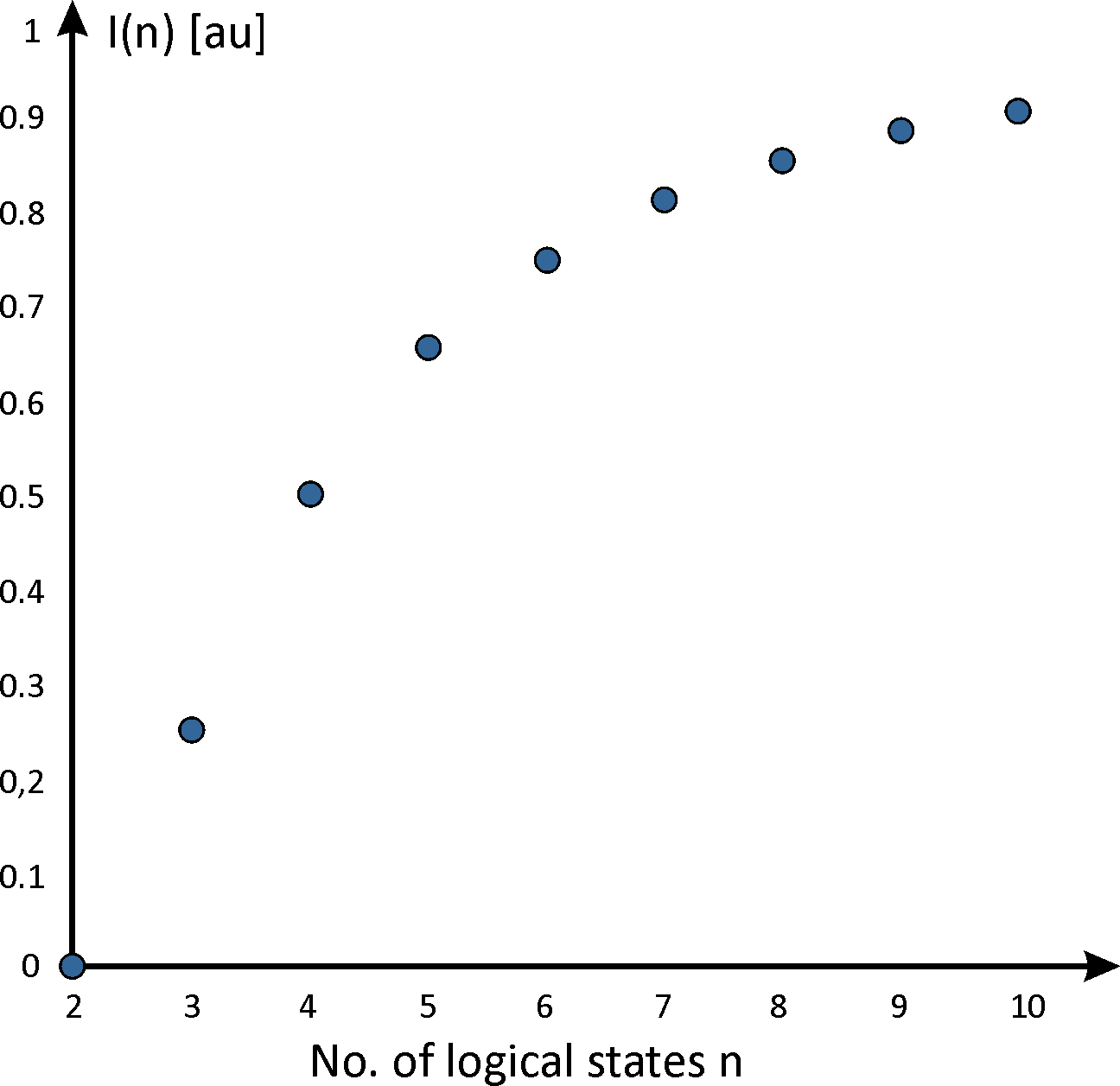}
\caption{The value of the boolean threshold for the chosen no. of logical states, determined by the particular numerical system}
\label{fig:threshold_analytics}
\end{figure}

Out of this, one can plot the dependence of the threshold for the subsequent numerical systems, which is shown in Figure \ref{fig:threshold_analytics}. This proves, that the protocol can operate with higher capacity, achieved by the higher density of values in subsequent numerical systems. 

The main purpose of the presented protocol is to assess the similarity between two signals without disclosing their actual content. The security level escalates with the number of logical states. In systems denser than binary, deducing one signal from the other becomes nontrivial even if the latter is known and a third party has information about the similarity of individual elements in both signals. However, denser numerical systems provide additional experimental challenges related to the possibility of differentiating between subsequent power levels, assuming that the boolean logic is considered and the threshold has to be set. To challenge the current approach to data computing, the following paragraphs present experimental results concerning the implementation of optical ternary reversible logic for comparing sets of signals with structured light.

\subsection*{Image comparison and signal encoding}

To present protocol capabilities, we determine the similarity of images, directly determining the Hamming distance, corresponding to the number of dissimilar elements in compared hash sequences. This, however, serves only as an exemplary scenario. The presented protocol is capable of comparing any set of signals, not limited to image comparison. As a source of signals, we chose a set of images consisting of the Cameraman's original image and its modified (distorted) versions. 

The efficient way of comparing sets of images requires the application of a fingerprinting algorithm, which instead of comparing large datasets will reduce the task to fingerprint comparison. For that purpose, we implemented the hash algorithm, where each image was converted into its hash representation using the open-source perceptual hash algorithm based on 2-D discrete cosine transform \cite{pHash}. The perceptual hashing algorithm can be reduced to just a few major steps. Firstly, the 2-D discrete cosine transform (D) of the input image $Img$ was calculated:


\begin{equation}
D = dct2(Img);
\end{equation}

Then, the size of the DCT was reduced, so only the low-frequency content was preserved.
\begin{equation}
D = D(1:q,1:q);
\end{equation}
Where $q$ determines the size of a hash, for the input $Img$ of 256x256 px, the $q=8$
This was followed by the computation of the global mean D value:

\begin{equation}
Av_{D} = \frac{1}{q^2} \sum_{i=1}^{q} \sum_{j=1}^{q} D(i, j)
\end{equation}

Where $i$ and $j$ denote indices of the matrix. 
Finally, the binary hash $H$ was constructed:
\begin{equation}
H(i, j) = \begin{cases} 1 & \text{if } D(i, j) > Av_D \\ 0 & \text{otherwise} \end{cases}
\end{equation}

The ternary hash $V$ can be deducted from its binary representation following the convention binary -> decimal -> ternary or by exchanging the last step of the algorithm to:
\begin{equation}
V(i, j) = \begin{cases} 2 & \text{if } D(i, j) > Av_D \\  1 & \text{if } D(i, j) = Av_D \\ 0 & \text{if } D(i, j) < Av_D \end{cases}
\end{equation}

\begin{figure}[ht]
\centering
\includegraphics[width=8cm]{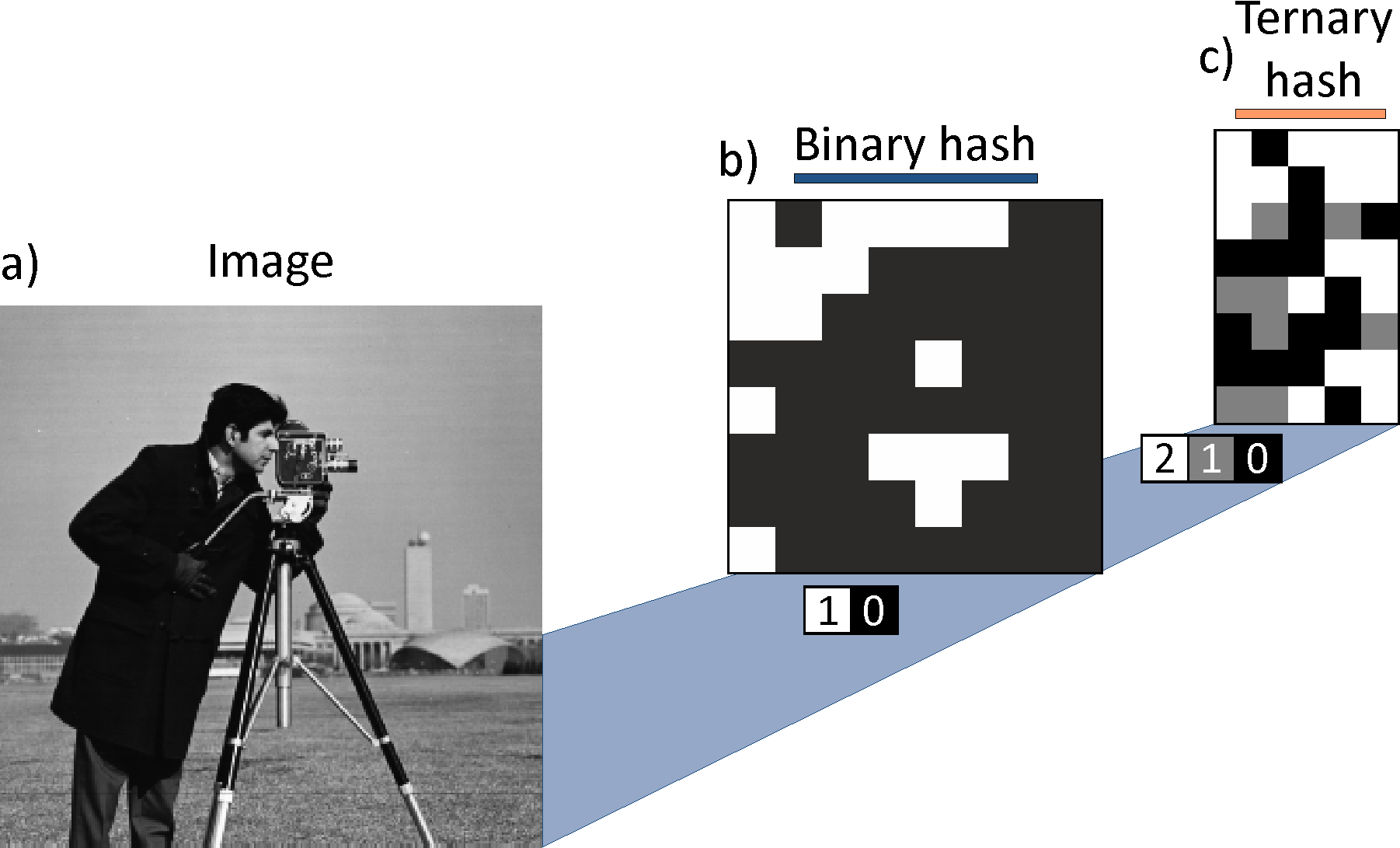}
\caption{The process of image hashing. a) Original image b) Its binary hash c) Image hash in a ternary basis. Blue and orange stripes were used as indicators of binary (blue) and ternary (orange) domains}
\label{fig:hash}
\end{figure}

Figure \ref{fig:hash} presents both the exemplary images and their hashes. To prepare the sequence that could be sent with the protocol, the hash matrix was transformed into a vector (signal), by row sectioning. We remind, that while we use numbers to represent logical values, they correspond to phase shift values, as shown in Figure \ref{fig:multilevel}.

In the subsequent step, each element of the created signal is encoded in the proper L-G mode and then two opposite L-G modes are superposed to make the final hologram. Therefore, the number of holograms corresponds to the no. of elements N in a particular hash sequence. While in this experiment we overlapped two L-G modes numerically, the same can be achieved through beam separation, where each L-G mode will be encoded separately to be later superposed, as in \cite{Szatkowski2021}. These two approaches share the same concept. 

While we decided to use the protocol to compare two images, the range of applications is much broader.  The next paragraph discusses the general message-passing protocol procedure, which can compare any signals, as long as they are represented in either binary or ternary logic.

\subsection*{Experimental setup} \label{sec:protocol}

\begin{figure}[ht]
\centering
\includegraphics[width=13cm]{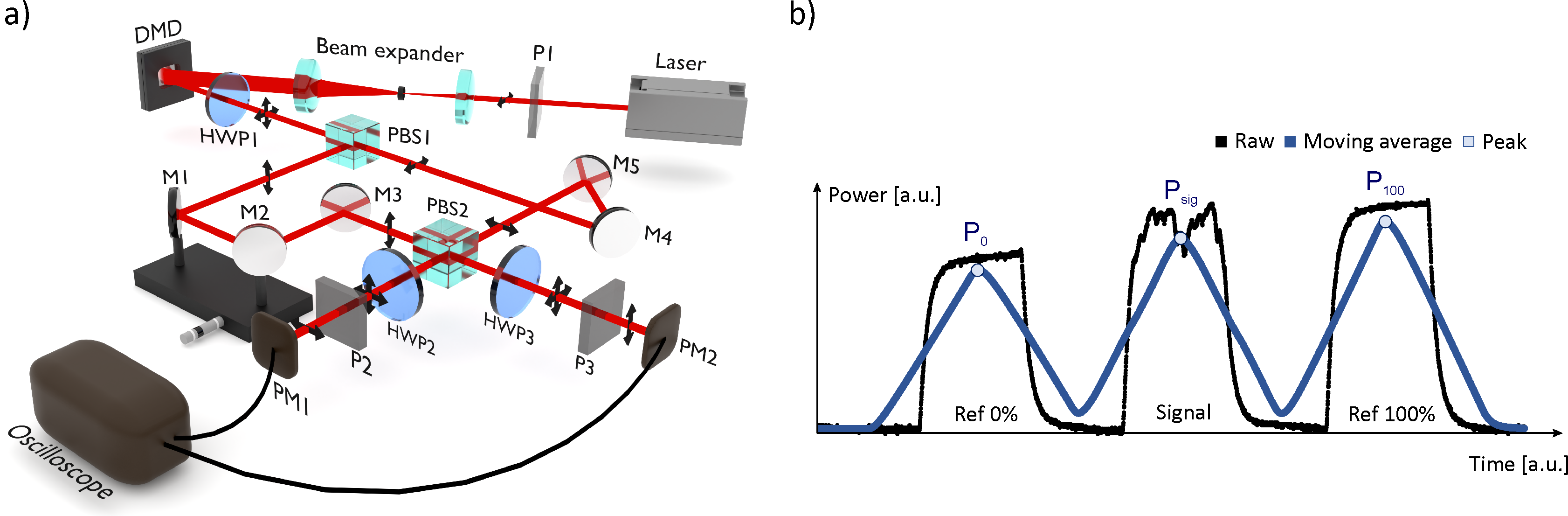}
\caption{a) The experimental setup b) The raw signal registered by the oscilloscope for a single detected packet (black), moving average calculated over the raw signal(blue), the detected peak values of the moving average (light blue)}
\label{fig:Setup}
\end{figure}

The experimental setup, shown in Figure \ref{fig:Setup}a, realizes the controlled-SWAP through the modified polarization Mach-Zehnder interferometer, with uneven no. of reflections in both arms (2 for the top and 5 for the bottom arm). However, the swap operation can also be achieved with other methods, including a tilted lens \cite{Luo2017} or diffraction grating \cite{Mohagheghian2023}.

The expanded laser beam with horizontal polarization illuminates the Digital Micromirror Device (DMD), which encodes the beam by displaying the superposition of two opposite L-G modes. Each possesses the signal encoded as their phase shift. The complex DMD phase-amplitude modulation has been achieved with the Lee method \cite{Lee1979, Conkey2012} and the final hologram has been displayed on the DMD. The reflected 1st order of diffraction was further propagated towards the interferometer. The half-wave plate (HWP1) acts as a first Hadamard gate and sets the diagonal polarization which is further split into horizontal and vertical counterparts by the polarization beam splitter (PBS1). The bottom arm of the interferometer, containing the vertical polarization component has an additional phase delay consisting of M1 and M2 mirrors mounted on the motorized stage, that controls the interference contrast at the detector. The bottom beam further propagates to the M3 mirror and reaches the second polarization beam splitter (PBS2). The upper beam experiences two reflections from M4 and M5 mirrors until it meets the bottom beam in the PBS2. After the PBS2 beams reach second Hadamard gates, realized through half-wave plates HWP2 and HWP3, respectively. At the last stage, orthogonally sets polarizers P1 and P2 implement vertical and horizontal polarization, respectively. The output signal is collected by the photodiode power sensor PM1 and PM2 which together with the sensor's console transmits the analog signal to the oscilloscope. The example dataset registered by the Oscilloscope for PM1 is presented in Figure \ref{fig:Setup}b. By incorporating local reference signals, we achieved the local normalization of the protocol response, which overcomes the air fluctuations affecting the measurement. Depending on the phase delay introduced by M1 and M2 mirrors, one can choose either PM1 or PM2 will perform the overlap analysis. As has been shown in \cite{Szatkowski2021}, both detectors are complementary, but one is enough to provide a sufficient analysis.

To determine the overlap OP between two sent messages, we detected the peak values, representing the mean power detected by the oscilloscope in the packet (Figure \ref{fig:Setup})b. In such a scenario, the overlap $OP$ of two signals can be calculated based on the mean signal value $P_{sig}$ in reference to $0\%$ and $100\%$ power values, $P_{0}$ and $P_{100}$, respectively:
\begin{equation}
OP=\frac{P_{sig}}{|P_{100}-P_{0}|}  
\end{equation}

This straightforwardly leads to the Hamming distance $HD$, which can be determined as:
\begin{equation}
HD=(100\%-OP)*N
\label{eq:HD}
\end{equation} 

Where N denotes the no. of elements in the hash sequence.

\subsection*{Experiment}

\begin{figure}[ht]
\centering
\includegraphics[width=12cm]{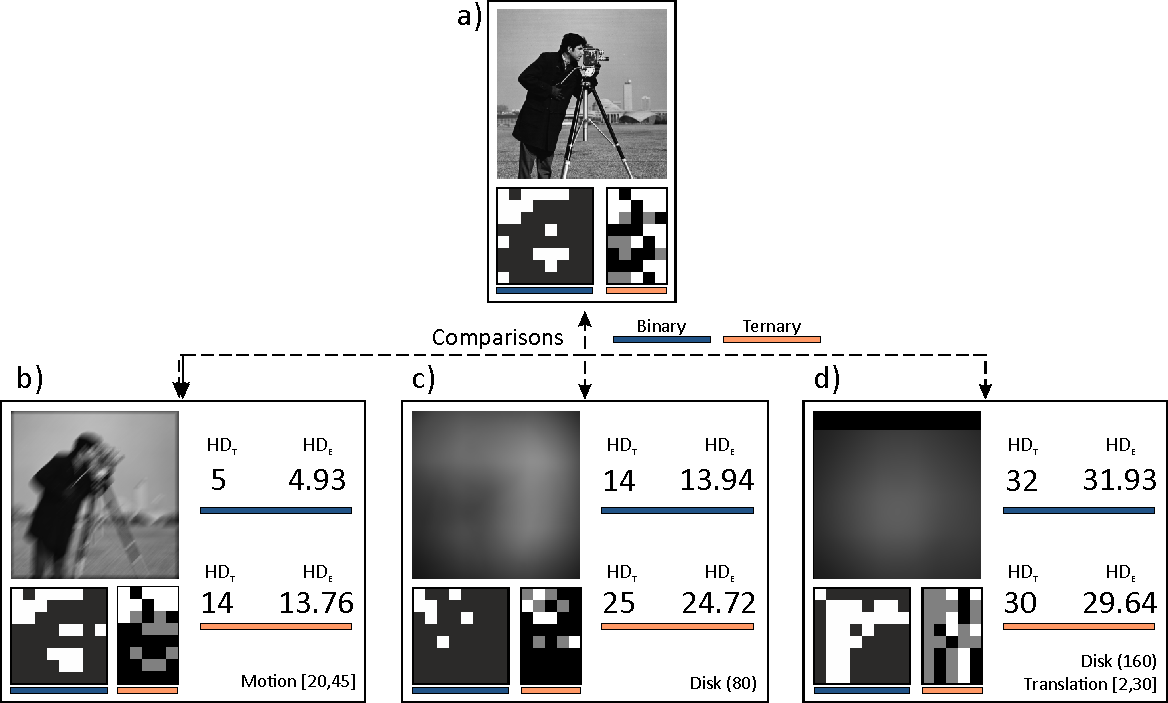}
\caption{Comparison of various images with their original counterpart. Bottom subfigures present a cluster, where each distorted image is followed by its hashes in binary (blue stripe) and ternary (orange stripe) basis. Each cluster presents the theoretical $HD_T$ and experimental $HD_E$ Hamming distance, for both binary (blue stripe) and ternary (orange stripe) signal comparison. a) The original b) distorted through the motion filter, c) disk filter, and d) disk filter and translation.}
\label{fig:results}
\end{figure}

To examine the protocol performance we compared a series of images consisting of an original Cameraman image and its distorted versions. Each time, the original image (Figure \ref{fig:results}a) was compared with its distorted counterpart (Figure \ref{fig:results}b-d), both represented as hash fingerprints in either binary (blue stripe) or ternary (orange strap) basis. Figure \ref{fig:results} presents both images, their hashes in each basis, as well as experimental results given by the Hamming distance (HD). 

The distortion was caused by the numerical image filtering and geometrical transforms. In particular, we applied the motion filter (Figure \ref{fig:results}b, disk filter (Figure \ref{fig:results}c), and disk filter followed by image translation (Figure \ref{fig:results}d) \cite{doi:https://doi.org/10.1002/9781119994398.ch2}. 

The theoretical HD $HD_T$ was calculated numerically, while the experimental HD $HD_E$ was determined through the equation \ref{eq:HD}. The length of signals was equal to N=64 for binary hash and N=40 for ternary hash. Each element of the signal has been displayed for 200 $\mu$s, corresponding to a 5kHz modulation frequency. This sets the value of the averaging filter, which was adjusted to match the length of a signal and started from the first non-0 element of Ref 0$\%$. In each case, the retrieved HD matched its expected value, when rounded to the closest integer. Because HD cannot be fractional, each time the experimentally retrieved $HD_E$ matches its theoretical value $HD_T$. 

Each of the free space communication systems, especially if based on interferometry,  is sensitive to the environmental conditions that may affect its stability. In particular, mechanical vibrations and air fluctuations may cause power variation at the detector. In the proposed solution, we overcome this issue by providing local references. Thus, in the given example, for a ternary basis, the similarity of two images is determined in less than 0.04s, without further post-processing. 

We performed the stability test, where reference signals Ref 0$\%$ and Ref $100\%$ were displayed one after another for the 120s, and the beam propagation distance across the setup did not exceed 3 meters. The standard deviation of the power difference between detected power values was less than 3$\%$, which in this example indicates that the system was stable for the period 3 000 times longer than required. The provided local reference algorithm made the protocol much less sensitive to environmental challenges.

\section{Methods}\label{sec11}
\subsection*{Message-passing protocol procedure}

The experimental setup presented in \ref{fig:Setup}a is built on the modified carrier frequency interferometer, which splits the incoming beam into separate polarization components. Therefore, in the aligning process, it is crucial to achieve an equal intensity after the PBS1 in each interferometric arm. This is controlled by rotating the HWP1 so that the incoming polarization will be diagonal and the PBS1 will split the beam in a 50:50 ratio. Then the aligning process does not differ, from the typical interferometer, however, for the proposed protocol it is crucial to control the optical path length difference with mirrors M1 and M2 mounted on the translational stage. The final criterion for determining if the proper optical path length difference between two interferometer arms is the power contrast between subsequent reference sets, observed by either PM1 or PM2 detector. The user would subsequently send two reference signals consisting of 0$\%$ and 100$\%$ of overlap adjusting the translational stage to achieve the highest power difference possible. We recommend the 100$\%$ difference to be at least 0.5 of the normalized power value.
If the alignment of the setup is performed. Then, the protocol procedure can be simplified into the following steps:

\begin{enumerate}
\item Determine the number N of elements (bits or trits) in the hash sequence and construct the reference signal sets (0$\%$ and 100$\%$ of overlap, respectively), having the same length as the hash sequence. 
\item Send the 0$\%$ overlap reference set continuously, without any delay between subsequent elements of the set. The set consists of all unequal elements leading to 0$\%$ of overlap between compared signals.
\item Follow the 0\% reference with the pause signal (no hologram displayed). Match the duration of the pause with the duration of the reference sequence.
\item Send both hash sequences that have to be compared. Each L-G mode carries the particular hash element of the original and distorted image accordingly.
\item Repeat the pause signal, as in the 3rd step.
\item Send the 100\% reference, similar to the 0\% reference signal in the 2nd step. This time comparing any signal with itself, to provide 100\% of overlap. 
\end{enumerate}

\section{Discussion}\label{sec12}

Implementation of reversible logic in the computational industry is considered one of the most promising solutions to lower the amount of energy per logical operation. In this work, we showed that the reversible logic can successfully be implemented with the structured light, proposing a novel communication protocol based on ternary phase-shifting encoding, where information is non-accessible at any point. This does not modify the encoding efficiency and straightforwardly increases the capacity of the protocol: the ternary basis requires fewer elements than binary to transmit the signal and the efficiency can be calculated as $E_f=(1-\frac{\log_2 3}{k})$ where $k$ denotes no. of elements in the signal represented in a binary basis. In practice, however, additional factors need to be considered, such as the efficiency of encoding and decoding data in a given numerical system, as well as the possibility of error correction in transmission. 

Apart from increasing the capacity, the security of the protocol has also been improved. One cannot directly decode one of the signals, even if both the outcome and the other signals are known. 

In the experimental example, we determined the Hamming Distance between pairs of images, represented by their hash fingerprints. The encoding procedure relies on phase-shifting and as long as the LG mode is created upfront, no further 2D phase modulation is required. Thus, it is possible to provide the external phase modulation, which should further speed up the encoding process, for instance using acousto- or electro- optic modulators.

This study has the potential to inspire quantum solutions with analogous architecture and can be practically applied in photonics circuits \cite{Ono2017}. The versatility of reversible logic in optical computing systems is highlighted by the ability to construct various logical gates using combinations of C-SWAP or CNOT gates, thanks to their universality \cite{nielsen00}. Integrating ternary logic into these systems positions them as excellent candidates to complement DNA data storage. This is particularly relevant as digital data-to-DNA conversion algorithms often require the use of ternary approaches \cite{Goldman2013}. 

The proposed communication protocol can be further developed. The potential directions of development are to combine the structured light phase-shifting encoding with the one where OAM is used as a logical value, breaking the extending the capacity limit \cite{Zhao2015}. Additionally, if the free space operating regime would be held, it is possible to spatially stack the protocol and simultaneously compare various sets of signals within the single optical system, at once.

\backmatter

\bmhead{Acknowledgments}

This research was funded in whole by National Science Centre, Poland, 2022/45/B/ST7/01234.

\section*{Declarations}

\begin{itemize}
\item Conflict of interest: The authors declare no conflicts of interest.
\item Authors' contributions: D.L.M. proposed the idea, M.S. and J.M. designed the experiment. K.M. developed the data analysis code, P.L. and J.W. built the experimental setup and analyzed the data, P.L. prepared the experimental setup figure. D.L.P and J.M. consulted the results and revised the manuscript, M.S. supervised the project, and prepared the manuscript. All authors discussed the results and the final version of the manuscript.
\item All the data that support the findings of this study are available from the
corresponding author on request.
\end{itemize}

\bibliography{sn-bibliography}

\end{document}